# Fermionic scenario for the destruction of superconductivity in ultrathin MoC films evidenced by STM measurements

P. Szabó,[1] T. Samuely,[1] V. Hašková,[1] J. Kačmarčík,[1] M. Žemlička,[2] M. Grajcar,[2]
J. G. Rodrigo,[3] and P. Samuely[1]

[1]*Centre of Ultra Low Temperature Physics @ Institute of Experimental Physics, Slovak Academy of Sciences and P. J. Šafárik University, SK-04001 Košice, Slovakia*
[2]*Department of Experimental Physics, Comenius University, SK-84248 Bratislava, Slovakia*
[3]*Laboratorio de Bajas Temperaturas, Departamento de Física de la Materia Condensada, Instituto de Ciencia de Materiales, Nicolás Cabrera, Condensed Matter Physics Center, Universidad Autónoma de Madrid, E-28049 Madrid, Spain*

We use sub-Kelvin scanning tunneling spectroscopy to investigate the suppression of superconductivity in homogeneously disordered ultrathin MoC films. We observe that the superconducting state remains spatially homogeneous even on the films of 3 nm thickness. The vortex imaging suggests the global phase coherence in our films. Upon decreasing thickness, when the superconducting transition drops from 8.5 to 1.2 K, the superconducting energy gap $\Delta$ follows perfectly $T_c$. All this is pointing to a two-stage fermionic scenario of the superconductor-insulator transition (SIT) via a metallic state as an alternative to the direct bosonic SIT scenario with a Cooper-pair insulating state evidenced by the last decade STM experiments.

**Keywords:** effect of disorder, superconducting films, STM, vortex phases

**PACS:** 74.62.En, 74.78.-w, 74.55.+v, 74.25.Uv

Superconductor-insulator transition (SIT) can be tuned in different ways, e.g. by increasing the physical or chemical disorder, by the change of charge-carrier density, by a magnetic field, etc. Decreasing the superconducting film thickness down to several atomic layers is yet another possibility [1]. Superconductivity is characterized by the order parameter $\Psi=\Delta\, e^{i\phi(r)}$, with the amplitude $\Delta$ and phase $\phi$. Two fundamental approaches describe SIT as a consequence of either the quasiparticle fluctuations affecting the amplitude $\Delta$ or the phase fluctuations of $\Psi$. The quasiparticle fluctuations correlate to the fermionic scenario [2,3]. There, disorder-enhanced Coulomb interaction breaks Cooper pairs into fermionic states leading to superconductor-(bad) metal transition (SMT) with $\Delta \rightarrow 0$ at $T_c \rightarrow 0$. At even higher disorder metal-insulator transition (MIT) follows due to Anderson localization. The bosonic scenario [4,5] assumes a direct SIT. On the superconducting side even in a homogeneously disordered system superconducting "puddles" with variant $\Delta$ emerge. In the insulating state Cooper pairs with a finite amplitude are still present, but their global phase coherence is lost. Comprehensive review on SIT as a quantum phase transition [6] resumes that many characteristics in both scenarios are very similar and probably only a probe directly sensitive to the local variations of the superconducting energy gap/order parameter can discern the realized mechanism. The scanning tunneling microscope (STM) is a unique probe with such a capability.

The available STM experiments on thin films of TiN, InO$_x$ and NbN [7-13] bring evidences that upon increasing disorder $\Delta$ decreases more slowly than $T_c$, the variation of $\Delta$ on a scale of the superconducting coherence length $\xi$ increases, a pseudogap appears above $T_c$ in the tunneling spectra, the spectral coherence peaks are suppressed, and the vortex lattice is fading out. This phenomenology strongly supports the bosonic scenario and raises the question about the universality of the bosonic SIT [14]. On the other hand, potential evidence for the fermionic mechanism was provided even sooner by the tunneling experiments on planar junctions on amorphous Bi and PbBi/Ge films [15,16] indicating that the Cooper pair amplitude vanishes at SIT. But the unambiguousness of these results was challenged by the fact that the spatially averaged tunneling spectra were gapless. This gaplessness could be explained by a very inhomogeneous gap distribution due to phase fluctuations. Then, superconducting pair correlations might also exist in insulators close to SIT, contradicting the fermionic scenario.

Here, we present sub-Kelvin STM experiments on ultra thin superconducting MoC films with atomic spatial resolution. We demonstrate the spatial homogeneity of the superconducting state for film thicknesses down to 3 nm and the closing of the superconducting energy gap/order parameter as $T_c$ vanishes, in agreement with the fermionic scenario. Thus, we bring the first direct evidence on the local behavior of the superconducting order parameter in this class of the transition.

The MoC films were prepared by the magnetron reactive sputtering from a Mo target in an argon-acetylene atmosphere onto a sapphire *c*-cut substrate. The details can be found elsewhere [17]. The preparation followed the procedure of Lee and Ketterson [18] who manufactured continuous MoC films down to 0.4 nm thickness showing the separatrix between the superconducting and insulating films at 1.3 nm thickness with a sheet resistance $R_S$ of about 2.8 kΩ.

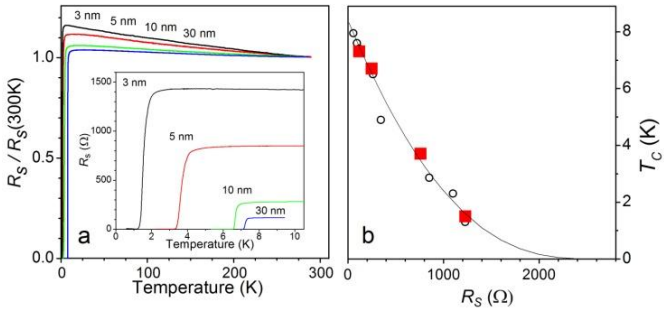

FIG. 1. (a) Temperature dependence of the sheet resistance $R_S$ normalized to its value at $T$ = 300 K for 3, 5, 10 and 30 nm MoC films. Inset: $R_s$ near $T_c$. (b)The critical temperature $T_c$ as a function of the sheet resistance $R_S$ for the measured films (squares) and other series (circles), solid line is Finkelstein's fit.

The inset of Fig. 1(a) shows the resistive transitions near $T_c$ on 30, 10, 5 and 3 nm thick MoC films. The gradual shift of $T_c$ (defined at 50% of $R_S$ above the transition) from 7.3 K to 6.7 K, 3.75 K and 1.2 K, resp. is accompanied by an increase of $R_S$ from several tens of Ohms to almost 1400 Ω. The transitions remain sharp while shifting to lower temperatures strongly suggesting that our films are homogeneously disordered [19]. The main part of Fig. 1(a) presents the overall temperature dependence of $R_S$ from room temperature down to the superconducting transition showing a negative derivative $dR_S/dT$ for all the films. Such a behavior can be described within the framework of the quantum corrections [20,21] to the standard Drude conductivity for disordered metals (weak localization and electron-electron interactions). These corrections are rather weak bringing about 15 % increase of $R_S$ before the thinnest 3 nm MoC film goes superconducting. Figure 1(b) shows a dependence of $T_c$ on $R_S$ (latter measured at 300 K) for the films used in the STM measurements (red squares) supplemented by another series of MoC films (circles). The solid line is a fit to the Finkelstein's model [3] indicating termination of superconductivity at $R_{cr} \approx 2.5$ kΩ close to the value determined in Ref.[18]. Magnetotransport and Hall-effect measurements allowed for determination of the superconducting coherence length $\xi$ which increases from 5 to 8 nm and the Ioffe-Regel product of $k_Fl$ dropping from 5 to 1.3 when the thickness of MoC film is diminished from 30 to 3 nm. Notably the carrier density is very high, $n \approx 10^{23}$ cm$^{-3}$, and not changing [22].

The scanning tunneling microscopy (STM) and spectroscopy (STS) experiments were performed by means of a sub-Kelvin STM system developed in Košice enabling measurements above $T$ = 280 mK and magnetic fields up to $B$ = 8 Tesla. Prior to STM experiments a protective layer on the surface of the films has been dissolved and the samples mounted within few minutes to the $^3$He cryostat. Surface topography has been done in the constant current mode. Conductance maps and vortex imaging have been studied via the Current Imaging Tunneling Spectroscopy mode [23] in a 128 x 128 grid.

The surface topography measurements have been realized on all our thin films at tunneling resistances in the range of 50 - 100 MΩ. The top panels of Fig. 2 show topographic images of 400 x 400 nm$^2$ surfaces obtained on 10, 5 and 3 nm films (a), (b) and (c), respectively at $T$ = 450 mK. Our films show clean surfaces, revealing compact polycrystalline structure with the lateral grain size of ~15-30 nm. Given the sharp superconducting transitions the grain boundaries are very transparent. An *rms* roughness of the films was typically 0.6-0.7 nm in agreement with the AFM topography [22]. A zoom of a 3 x 2.5 nm$^2$ area is displayed in the topography of the 5 nm film (inset in the topography image), showing atomically resolved surfaces with a distorted hexagonal crystal symmetry [24,25], which allow for correlated topographic (STM) and spectroscopic (STS) studies. It also proves that grains are single nanocrystals.

The tunneling conductance curves $G(V,x,y)=dI(V,x,y)/dV$ vs. $V$ are calculated by numerical differentiation of the locally measured $I – V$ characteristics. Since the metallic Au tip features a constant density of states at the measured bias energies, each of the differential conductance vs. voltage spectra reflects the superconducting density of states (SDOS) of the MoC films, smeared by ~ ± $2k_BT$ in energy at the respective temperature. Consequently, in the low temperature limit, the differential conductance measures directly the SDOS. The 3D plots shown in the middle panels of Fig. 2 display three series of 100 tunneling spectra $G(V,x,y)$ taken along 200-250 nm line at $T$ = 450 mK on the 10, 5, and 3 nm MoC films. The spectra are normalized to the conductance values $G_N$ at $eV$ = 4$\Delta$ above the gap energies. In the 10 nm films, the spectra show a pronounced superconducting structure, with well-defined coherent peaks near the gap energies at $V \sim \pm \Delta/e$ without any significant spatial variation across the sample surface. Notably, we always observed a finite conductance inside the gap region, with a value at zero bias, $G(V = 0,x,y)$, of around $0.1G_N$. This is not compatible with a pure BCS SDOS considering the thermal smearing at 450 mK. We address this phenomenon below but note that the complex conductivity derived from our recent experimental transmission measurements on coplanar waveguide resonators made of 10 nm MoC film shows a finite quasiparticle lifetime consistent with the existence in-gap quasiparticle states [26].

The series taken on the 5 nm film reveals gap-like features with coherent peaks located at a smaller bias (mean value of the peak position is $V$ = 0.88 mV with σ = 0.022 mV) as expected for a sample with lower $T_c$, but their height is significantly smaller than in the case of the 10 nm film Moreover, the intensity of the in-gap states is substantially increased to a value of $G(V = 0,x,y)$ around $0.3G_N$. This tendency continues on the 3 nm MoC film where the series of spectra features a reduced tunneling conductance around the zero-bias voltage and the in-gap states bring the value of $G(V = 0,x,y)$ to around $0.7G_N$ and almost no signatures of coherence peaks. Moreover, a quasi-linearly increasing background of the tunneling conductance at higher bias voltages is present. The latter finding indicates an appearance of the Altshuler-Aronov effect of suppression of density of states due to a disorder-enhanced Coulomb interaction [20].

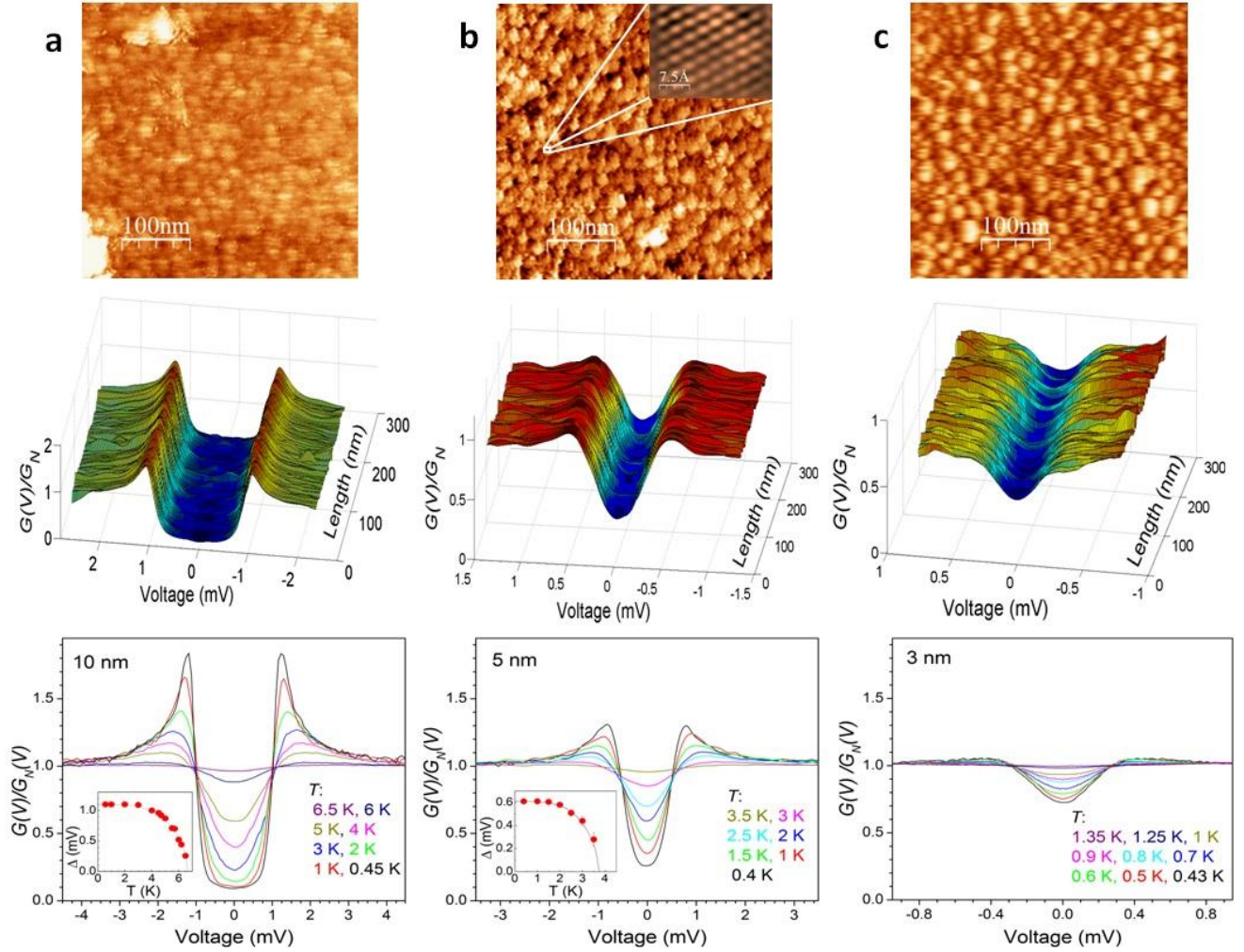

FIG. 2. STM surface topography and locally measured tunneling spectra for 10, 5 and 3 nm MoC films (a), (b) and (c), resp. Top: STM surface topographies at 500 mK, zoom in 5 nm film shows atomic structure. Middle: 100 STS spectra along ~ 200 nm line on the surface of the respective film taken at 450 mK. Bottom: Temperature dependence of a typical tunneling spectrum at indicated temperatures. Inset of (a) and (b): Temperature dependence of the gap determined from fits to Dynes formula (points) and BCS-like $\Delta(T)$ dependence (line).

For the different particular film thicknesses we analyzed numerous series of spectra like the ones presented in Fig. 2. None of them show any characteristic length-scale of variations which would indicate emerging superconducting granularity like e.g. in TiN [7]. On the contrary, the same variations are found independently on a bias voltage for spectra taken along lines with a step size of 2 nm (Fig. 2) or e.g. 0.05 nm (not shown). In the conductance maps (recorded at zero bias or at biases close to the gap value) on all studied MoC films, a variation with the same standard deviation of about 0.05 in normalized conductance is found. As such, these variations result from the noise of the apparatus rather than physical properties of the superconducting films. This implies a high spatial homogeneity of superconductivity in MoC.

Figure 3 (a-c) presents a surface analysis performed on another 5 nm MoC film. In (a) the STM surface topography of a small area of 95 x 65 nm$^2$ shows a protruding grain (white spot). Figure 3(b) displays the gap map (bias voltage of coherence peaks) of the same area taken at $T = 450$ mK and (c) provides the topography (red curve - left scale) and the gap map (blue curve – right scale) along the gray lines in (a) and (b). On the grain where the film is thicker a larger gap is found, notably in an area of size much larger than the superconducting coherence length, $\xi \approx 8$ nm. From these data it is evident that the only systematic change of the gap value in our MoC films is due to the change of their thickness and that the superconducting state remains spatially homogeneous on all MoC films with constant thickness. This is very different from the STS measurements on TiN, InO$_x$ and NbN thin films showing an emergent granularity in the superconducting condensate upon increasing disorder on the scale of the coherence length [7,9,12,13].

Fig. 3(d) shows a zero-bias conductance map obtained on the 5 nm film on a flat surface of 220 x 190 nm$^2$ at $T = 450$ mK in a magnetic field of $B = 1$ T. A distorted vortex lattice is observed. At 1 T, the average intervortex distance for the triangular Abrikosov lattice, $d_v = 1.075 \, (\Phi_0/B)^{1/2}$ is about 50 nm, resulting in some 21 vortices for an area equivalent to the one in the figure, as we actually find in our measurement. We have observed the presence of vortices in MoC films down to 5 nm thickness. In the case of 3 nm the low contrast between the superconducting and normal spectra prevents vortex imaging. The presence of vortices in MoC brings evidence for a global phase coherence in the samples allowing for supercurrent flow around the vortex cores.

The bottom panels of Fig. 2 show the effect of temperature on the tunneling spectra normalized to the tunneling curves $G_N(V)$ measured in the normal state above $T_c$. $G_N(V)$ of the 10 nm and 5 nm films were constant in the presented bias voltage range while on the 3 nm film a weak V-shaped normal state is present fitting the background shown in Fig. 2(c), middle panel. The local $T_c$'s have been established from the STS data as the temperatures where superconducting features disappear. Then, $T_c$'s are 6.7 K and 3.7 K and 1.2-1.25 K for the 10, 5 and 3 nm films, resp. in agreement with the transport data (see Fig. 1). No indications of a pseudogap featuring a reduced DOS above the bulk $T_c$ have been observed. We have measured 5-10 temperature dependencies on all our films taken at different locations but no variations of the local $T_c$ on a particular film have been observed. It again supports the spatial homogeneity of the superconductivity of the MoC films.

To fit our spectra with finite in-gap states the Dynes modification of the BCS density of states $N(E) = Re\{E/(E-\Delta)^{(1/2)}\}$, where the complex energy $E = E' + i\Gamma$ with a smearing parameter $\Gamma$ has been taken into account [27]. The temperature dependencies of $\Delta$ have been determined from fits to the thermally smeared Dynes formula on the 10 and 5 nm films. The large smearing effect in the 3 nm film prevented reliable fits at higher temperatures. The resulting $\Delta(T)$ dependencies presented in the insets of the bottom panels of Fig. 2 follow the predictions of the BCS theory (solid lines). In the fits $\Gamma$ is found to be temperature independent up to ~ $0.5T_c$, while at higher temperatures is increased up to ~ $4\Gamma(0)$ at $T_c$.

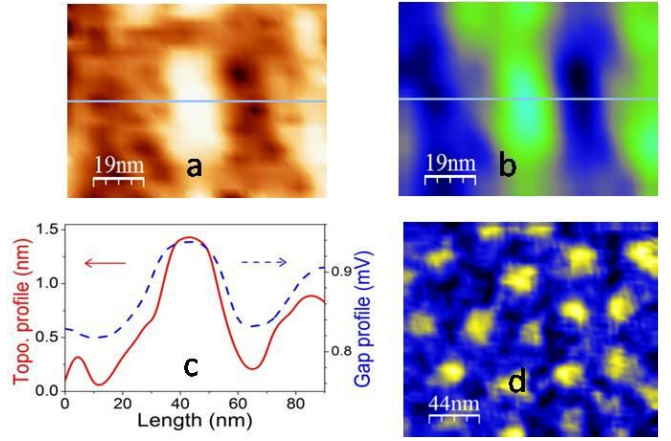

FIG. 3 (a) STM topography on 5 nm MoC film showing protruding grain as a white spot. (b) Gap map of the same area taken at $T = 450$ mK. (c) Profiles of topography (full curve) and gap map (dashed curve) along the gray lines in (a) and (b). (d) Vortex image on a flat surface of 5 nm MoC film at 450 mK in $B = 1$ T showing 21 vortices.

The tunneling spectra at $T = 450$ mK (solid lines) together with the Dynes fits (open symbols) are resumed in Fig. 4(a). There we have also included the result for the 30 nm thick MoC film which displays a hard spectral gap with a flat zero tunneling conductance in a finite range of bias voltages inside the gap region. The fit provides $\Delta(0) = 1.23$ meV with $\Gamma = 0$ fully in line with the BCS theory. The observation of the hard gap here proves that the spectral smearing on the other samples is not due to a lack of energy resolution. As sample thickness is decreased, the inferred gap/order parameter $\Delta(0)$ is decreased in a sequence $\Delta(0) = 1.1$, 0.62 and 0.2 meV for the 10, 5 and 3 nm film thicknesses, respectively. As presented in Fig. 4 (b) the thickness dependence of the superconducting gap $\Delta$ (red squares) perfectly follows the thickness dependence of the transition temperature $T_c$ (circles). Figure 4(c) shows that the superconducting coupling ratio $2\Delta/k_BT_c$ remains constant and being equal to about 3.8 for all the studied MoC films. All this strongly suggests that superconductivity is suppressed ($T_c \to 0$) when $\Delta \to 0$.

We can now compare our findings with the bosonic SIT "phenomenology" found in the STS experiments on TiN, InO$_x$ and NbN films [7-13]. In contrast to those experiments we have found that upon increasing disorder $\Delta$ decreases in the same way as $T_c$, there are no spatial variations of $\Delta$ and no pseudogap above $T_c$. Also, the vortex lattice is present. In the strongly disordered MoC films the spectral coherence peaks are suppressed, but in a different way than for example in InO$_x$ [10]. In MoC their suppression is correlated with the increase of the in-gap states which can be quantitatively accounted for by the increasing relative strength of the Dynes parameter normalized to the zero-temperature gap value, $\Gamma/\Delta$. Figure

4(c) shows a steep increase of this parameter from zero (30 nm film) to 0.9 for the 3 nm film.

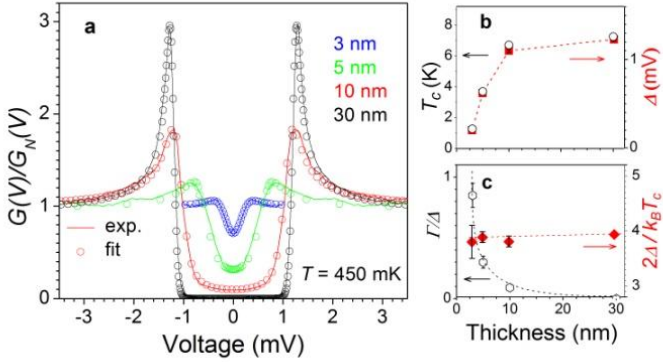

FIG. 4. (a) Typical tunneling conductances of the 3, 5, 10 and 30 nm thin MoC films at $T = 450$ mK – solid lines. The symbols are fits to the thermally smeared Dynes formula. b) Film thickness dependence of the superconducting energy gap (red solid symbols – right scale) and of the critical temperature determined from tunneling experiment (open circles – left scale). (c) Film thickness dependence of the Dynes smearing parameter normalized to the gap - $\Gamma/\Delta$ at $T = 450$ mK (open circles - left scale) and of the superconducting coupling strength $2\Delta/k_BT_c$ (red solid diamonds – right scale).

As mentioned above, broadened SDOS have already been observed on other films near SIT. Valles *et al.* found heavily broadened SDOS in planar tunnel junctions made on ultrathin Bi film with $T_c = 0.7$ K [15] but also on the series of ultrathin PbBi/Ge films [16], where it was attributed to spatial variations of the amplitude of the order parameter. In the PbBi films the broadening increases as $T_c$ decreases with maximum $\Gamma/\Delta \sim 0.4$ for $T_c = 0.96$ K. Our spatially homogeneous spectra clearly demonstrate that the origin of the broadening is not the uneven distribution of the gap amplitude. Hence, although the source of the in-gap states remains unclear [28], our findings unequivocally corroborate the fermionic scenario.

The recent measurements of the Little-Park oscillations in the magnetoresistance on uniform a-Bi films with nanohoneycomb array of holes also support the fermionic SIT [29]. Then, a question arises what physical parameter decides on the insulating ground state upon SIT, fermionic or Cooper-pair insulating one. The MoC superconductors differ from TiN, InO$_x$ and NbN for example by the fact that the quantum corrections to the resistivity are small and the charge carrier density is high and not changing upon increasing disorder. If these are important ingredients for fermionic scenario, it remains for further studies.

In conclusion, STM and STS studies on ultrathin MoC films provide evidence that, in contrast to TiN, InO$_x$ and NbN, where the bosonic scenario of SIT is found, in the ultrathin MoC films the superconducting energy gap or order parameter terminates, ($\Delta \to 0$) as the bulk superconductivity ceases with $T_c \to 0$, the global superconducting coherence in MoC films is manifested by the presence of superconducting vortices and most importantly, the superconducting state is very homogeneous for all the thicknesses down to 3 nm where the superconducting transition is suppressed from bulk $T_c = 8.5$ K to 1.3 K and the strong disorder is characterized by $k_Fl$ close to unity. All these observations point to the fermionic route of the SIT confirming that there are at least two different scenarios of SIT that can be realized depending on the physical parameters of the systems.

We gratefully acknowledge helpful conversations with M. Skvortsov, L. Ioffe, D. Roditchev, H. Suderow, and R. Hlubina. This work was supported by the projects APVV-14-0605, APVV-0036-11, VEGA 2/0135/13, VEGA 1/0409/15, EU ERDF-ITMS 26220120005, EU ERDF-ITMS 26220120047 and the COST action MP1201 as well as by the U.S. Steel Košice, s.r.o. J.G.R. also acknowledges support from projects FIS2014-54498-R (MINECO, Spain) and P2013/MIT-3007 MAD-2D (Comunidad de Madrid, Spain).